\documentclass[usegraphicx,usenatbib]{mn2e}

\usepackage{url}

\bibpunct{(}{)}{;}{a}{}{,}
\usepackage{aasmacros}

\newcommand{\simlt}{\mathrel{\hbox{\rlap{\hbox{\lower4pt\hbox{$\sim$}}}\hbox{$<$}}}}
\newcommand{\simgt}{\mathrel{\hbox{\rlap{\hbox{\lower4pt\hbox{$\sim$}}}\hbox{$>$}}}}

\title[Superluminous Supernovae Rates]{Rates of Superluminous Supernovae at $z\sim0.2$}

\author[Quimby et al.]{
\parbox[t]{\textwidth}{
  Robert M. Quimby$^{1}$\thanks{E-mail: robert.quimby@ipmu.jp},
  Fang Yuan$^{2}$,
  Carl Akerlof$^{3}$,
  and
  J. Craig Wheeler$^{4}$
}
\vspace*{6pt}\\
$^{1}${\it 
  Kavli IPMU, University of Tokyo, 
  5-1-5 Kashiwanoha, 
  Kashiwa-shi, Chiba, 277-8583, Japan}
\\
$^{2}${\it 
    Research School of Astronomy and Astrophysics, 
    The Australian National University,
    Weston Creek, ACT 2611, Australia}
\\
$^{3}${\it
  Physics Department, 
  University of Michigan, 
  Ann Arbor, MI 48109, USA}
\\
$^{4}${\it
  Department of Astronomy, McDonald Observatory, 
  University of Texas, 
  Austin, TX 78712}
}

\date{Accepted ---. Received ---}
\volume{000}
\pagerange{1\,-\,??} \pubyear{---}
\begin{document}
\maketitle

\label{firstpage}

\begin{abstract}
We calculate the volumetric rate of superluminous supernovae (SLSNe)
based on 5 events discovered with the ROTSE-IIIb telescope. We gather
light curves of 19 events from the literature and our own unpublished
data and employ crude k-corrections to constrain the pseudo-absolute
magnitude distributions in the rest frame ROTSE-IIIb (unfiltered) band
pass for both the hydrogen poor (SLSN-I) and hydrogen rich (SLSN-II)
populations. We find that the peak magnitudes of the available SLSN-I
are narrowly distributed ($M = -21.7 \pm 0.4$) in our unfiltered band
pass and may suggest an even tighter intrinsic distribution when the
effects of dust are considered, although the sample may be skewed by
selection and publication biases. The presence of OII features near
maximum light may uniquely signal a high luminosity event, and we
suggest further observational and theoretical work is warranted to
assess the possible utility of such SN\,2005ap-like SLSN-I as distance
indicators. Using the pseudo-absolute magnitude distributions derived
from the light curve sample, we measure the SLSN-I rate to be about
$(32^{+77}_{-26})$\,events\,Gpc$^{-3}$\,yr$^{-1}$\,$h_{71}^{3}$ at a
weighted redshift of $\overline{z} = 0.17$, and the SLSN-II rate to be
about
$(151^{+151}_{-82})$\,events\,Gpc$^{-3}$\,yr$^{-1}$\,$h_{71}^{3}$ at
$\overline{z} = 0.15$. Given that the exact nature and limits of these
populations are still unknown, we discuss how it may be difficult to
distinguish these rare SLSNe from other transient phenomena such as
AGN activity and tidal disruption events even when multi-band
photometry, spectroscopy, or even high resolution imaging are
available. Including one spectroscopically peculiar event, we
determine a total rate for SLSN-like events of
$(199^{+137}_{-86})$\,events\,Gpc$^{-3}$\,yr$^{-1}$\,$h_{71}^{3}$ at
$\overline{z} = 0.16$.

\end{abstract}


\section{Introduction}

There is a growing sample of supernovae with peak luminosities over 30
times brighter than average (based on $\overline{M} = -16.8$ from the
volume limited Lick Observatory Supernova Search,
\citealt{li2011a}). These superluminous supernovae (SLSNe), were not
identified in the first 60 years of studies following the pioneering
work of Zwicky and Baade, which marked the beginning of systematic
searches for extragalactic transients \citep{zwicky1938}.

This lack in what would naively seem to be the easiest kind of
supernova to discover could in part be explained if the historical
searches were simply looking in the wrong place; indeed, the sample of
SLSNe published thus far shows a preference for low luminosity
galaxies \citep{neill2011} and possibly galaxy cores--two environments
neglected by early surveys. But the greater distance over which the
SLSNe are visible further suggests that the intrinsic rates must be
low lest they be discovered in the background of a targeted galaxy.

The origin of these events is still a matter of debate (for a recent
review, see \citealt{galyam2012}). While the host environments and
energetics may best be understood in the context of massive stellar
explosions from young, star-forming environments, the ultimate power
source is less clear. The decay of radioactive $^{56}$Ni, the source
of a Type Ia explosion's brilliance, may be involved in certain events
(\citealt{galyam2009}, but see also \citealt{dessart2012}), but for
others, it cannot be more than a minor contributor
\citep[e.g.][]{pastorello2010,quimby2011,chomiuk2011}. Some events
show narrow emission lines that indicate that an interaction with a
slow-moving CSM may drain the large store of kinetic energy from the
supernova ejecta and transfer this into radiated light
\citep[e.g.][]{smith2007,ofek2007,smith2008,drake2010}. Others show no
such evidence for a long-lived interaction
\citep[e.g.][]{miller2009,gezari2009}. Some SLSNe may draw their power
from a compact object that forms in the collapse of the progenitor
star \citep[e.g.][]{kasen_bildsten2010,woosley2010,ouyed2012}.

Whatever their origin, the high luminosities of SLSNe make them
detectable from large distances, and they are thus capable of
shepherding information from the early universe to the
present. Intervening clouds of gas, including gas in the vicinity of
the progenitor, can leave absorption signatures on the spectra of
SLSNe, and these may transmit the chemical composition of these
otherwise undetectable systems \citep{quimby2011,berger2012}. If the
SLSNe are, in fact, products of massive stars, then their volumetric
rates will be entwined with the cosmic star-formation history and the
initial mass function \citep{tanaka2012}. Finally, if the luminosities
fall within a narrow enough range, or if they can be predicted from
secondary indicators such as the light curve width, then SLSNe may
serve as standard candles.

Among the first clear SLSN discoveries are several contributions from
the ROTSE-IIIb telescope at the McDonald Observatory. Although modest
in size (the primary mirror is just 0.45\,m), ROTSE-IIIb has a wide
field of view (about $1.85 \times 1.85$ degrees per exposure), which
allows large swaths of sky to be monitored \citep{akerlof2003}. As the
discoveries are necessarily bright ($M \simlt 18.3$), follow-up
spectroscopy is relatively inexpensive.

ROTSE-IIIb detected one of the first SLSNe, SN\,2006gy
\citep{smith2007}. Classified as a Type\,IIn from the narrow hydrogen
emission lines apparent in its spectra
\citep{harutyunyan2006,prieto2006,foley2006}, SN\,2006gy shone
brighter than a typical Type\,Ia supernova at peak for 5 months and
radiated over $10^{51}$\,erg in optical light alone. Other SLSN-II
discoveries have followed including SN\,2003ma \citep{rest2011},
SN\,2008am \citep{chatzopoulos2011}, and SN\,2008fz
\citep{drake2010}. Interaction with circum-stellar media clearly plays
an important role in these events and may be responsible for their
extreme luminosities \citep[e.g.][]{smith_mccray2007, smith2008b,
  chevalier_irwin2011}.

The nature of a prior ROTSE-IIIb discovery, SN\,2005ap, was not
immediately apparent. The first spectrum lacked the strong P-Cygni
profiles typically seen in supernova spectra, but the rise in flux
toward shorter wavelengths did indicate at hot photosphere
\citep{quimby2007c}. Later discoveries by the Palomar Transient
Factory (PTF; \citealt{law2009,rau2009}) and Pan-STARRS
\citep{kaiser2010} of similar objects show that this class of objects
is depleted in hydrogen and thus cannot produce its luminosity through
interactions with a hydrogen rich CSM
\citep{pastorello2010,quimby2011,chomiuk2011}.

A rough estimation of the rate of objects similar to SN\,2005ap was
presented in \citet{quimby2008}. The derivation used the discovery
ratio of SN\,2005ap to Type\,Ia supernovae in the ROTSE-IIIb sample
(at that time) and a simplistic estimate of the search volume ratio
for these to estimate a rate relative to the Type\,Ia
rate{\footnote{We note that the volume ratio published in
    \citet{quimby2008} is erroneous. Given the stated procedure, the
    calculation should have employed a volume ratio of about 27, not
    3.}}.  \citet{miller2009} followed this method with a larger
sample from ROTSE-IIIb to derive an approximate rate for objects
similar to SN\,2005ap and SN\,2008es relative to the Type\,Ia
rate. They find about 1 such event for every 350 Type Ia. They further
note that the non-detection of such events from the KAIT SN search
\citep{filippenko2001} implies that the rate in large galaxies is less
than 1/160 times the local Type II rate (note that 2006gy was actually
detected by the KAIT SN search in such a host, but it was not
discovered due to a selection bias). Recently, \citet{cooke2012}
estimated the high redshift SLSN rate using two transients found at
$z=2.05$ and $z=3.90$ in archival SNLS data. These were not
spectroscopically confirmed, but they imply a SLSN rate at $z=2-4$ of
$\sim 400$\,Gpc$^{-3}$\,yr$^{-1}$\,$h_{71}^{3}$.

\citet{quimby2012}, have measured the volumetric rate of Type\,Ia
supernovae using background discoveries from the ROTSE-IIIb
searches. In this paper, we measure the rates of SLSNe from the same
survey over the same period. It is not yet clear if SLSNe are produced
from one or multiple channels, so we will supply the necessary
parameters for the reader to determine the approximate rates (or
limits) of individual events from our survey, or to group events into
physically related groups. We also supply rate measurements of the bulk
SLSN-I and SLSN-II populations based on 1 and 3 discoveries,
respectively. We further supply the rate for an all inclusive category
of SLSN-like events, which includes one additional (as yet unverified)
SLSN candidate in addition to the confirmed SLSN-I and SLSN-II.  We
discuss this event and the full sample in more detail in
\S\ref{sample}.

These rates may be compared against prospective progenitor rates
(including any high luminosity tail of normal supernova) to help
constrain the origin of these events. The rates of Type Ia supernovae
and, in particular, core collapse events spanning the Type IIn through
Type Ic classes, have previously been used to make such inferences and
continue to be useful for such studies (see \citealt{graur2011},
\citealt{horiuchi2011}, \citealt{dahlen2012}, and references therein).

For our rate estimation, we perform a Monte Carlo simulation where the
light curves of simulated SLSNe are compared against observations
logged by ROTSE-IIIb to estimate the fraction of events we would
recover over a variety of distances. We construct a sample of light
curve templates from published SLSNe in \S\ref{lc} and use the peak
magnitudes from these to estimate the pseudo-absolute magnitude
distribution (pAMD), which is the intrinsic luminosity function mixed
with the host absorption distribution, in \S\ref{pamd}. The rates are
presented in \S\ref{rates}, and discussion and conclusions are given
in \S\ref{conclusions}.

\begin{table*}
\centering
\caption{Background SLSN-like Events Discovered with ROTSE-IIIb by Feb. 1, 2009.}
\label{table:disco}
\begin{tabular}{lrllllll}
\hline
Name & Disc. Date & RA & Dec & z & Type & Reference \\
\hline
2005ap & Mar  3, 2005 & 13:01:14.8 & +27:43:31 &  0.283 &   Ic & \citet{quimby2007c}\\
2006tf & Dec 12, 2006 & 12:46:15.8 & +11:25:56 &  0.074 &  IIn & \citet{smith2008}\\
2008am & Jan 10, 2008 & 12:28:36.3 & +15:34:49 &  0.234 &  IIn & \citet{chatzopoulos2011}\\
2008es & Apr 26, 2008 & 11:56:49.1 & +54:27:26 &  0.202 &   II & \citet{gezari2009}\\
Dougie & Jan 21, 2009 & 12:08:47.9 & +43:01:21 &  0.191 &    ? & Vinko et al. (in prep.)\\
\hline
\end{tabular}
\end{table*}

\section{The ROTSE-IIIb SLSN Sample}\label{sample}

Our sample is drawn from two similar surveys conducted with the
ROTSE-IIIb telescope: The Texas Supernova Search (TSS;
\citealt{quimby_phd}) and the ROTSE Supernova Verification Project
(RSVP; \citealt{yuan_phd}). Both surveys used ROTSE-IIIb's wide field
of view to canvas about 500 square degrees on the sky with preference
given to areas with high concentrations of galaxies in the local
($D<200$\,Mpc) universe.

Between November 1, 2004 and January 31, 2009, ROTSE-IIIb discovered
76 supernovae. All of these were spectroscopically confirmed by us or
others in the community. The highest luminosity Type\,Ia found in our
survey is SN\,2007if at about $M_r=-20.4$\,mag
\citep{scalzo2010,yuan2010}. Here we consider only higher luminosity
events, so we remove SN\,2007if and 69 fainter objects. To calculate a
volumetric rate that is unbiased by the local density enhancements
(e.g. galaxy clusters) targeted by our search, we must further remove
discoveries found about 50\,Mpc behind the targeted objects or
less. This cut eliminates what is perhaps the best known SLSN
discovery from ROTSE-IIIb: SN\,2006gy
\citep[][]{smith2007,ofek2007}. This object was found in a massive
galaxy (NGC\,1260) residing in the Perseus Galaxy Cluster, which we
had specifically targeted.

After these cuts, we are left with the five events listed in
table\,\ref{table:disco}. SN\,2005ap \citep{quimby2007c} was first
detected before SN\,2006gy, but it could not be identified as a SLSN
until after the discovery of SN\,2006gy reset the accepted limits of
the supernova peak absolute magnitude distribution. SN\,2005ap is the
first member of the hydrogen-poor, SLSN-I group. SN\,2006tf
\citep{smith2008} and SN\,2008am \citep{chatzopoulos2011} both show
hydrogen Balmer lines of relatively narrow widths in their spectra (in
addition to broader components), and can be grouped in the SLSN-II
category. The hydrogen features seen in the spectra of SN\,2008es
\citep{miller2009,gezari2009} lack these narrow emission peaks, and it
was not until the initially hot, blue continuum cooled and faded after
peak that broad P-Cygni profiles clearly emerged. This points to some
differences in the progenitor system--but not necessarily profound
differences \citep[cf.][]{moriya_tominaga2012}, and SN\,2008es can be
grouped with the other SLSN-II based on the eventual signs of
hydrogen. Finally, ROTSE-IIIb detected a high luminosity event
internally designated as ``Dougie'' (sometimes called
ROTSE3\,J120847.9+430121). Dougie exhibited a mostly featureless blue
continuum, which was seen to redden over time; however, the broad
features typical of supernovae were never observed. It is possible
that Dougie is a tidal disruption event or an unusual AGN outburst
(although no x-ray emission was detected), but we do not rule out a
connection to SLSNe at this time. Full details of this event will be
presented later (Vinko et al. in prep.).

In summary, we have just one SLSN-I (SN\,2005ap), three SLSN-II (SNe
2006tf, 2008es, and 2008am), and one additional, SLSN-like event
(Dougie) to use in our rate calculations. If Dougie is a supernova,
its lack of spectroscopic evidence for hydrogen would place it in the
SLSN-I group, and we will supply SLSN-I rates with and without this
event.

\section{Light Curves and pseudo-Absolute Magnitude Distributions of SLSNe}\label{lcsample}

A key factor in determining the rates of supernovae is accurately
knowing the peak magnitude distribution with the effects of host
absorption factored in. Here we devise representative light curves and
pseudo-absolute magnitude distribution (pAMD) models in our unfiltered
band pass for both SLSNe-I and SLSNe-II that can be used to determine
our survey efficiency and thus rates. The pAMDs required for our rate
study give the distribution of absolute magnitudes, without correction
for absorption external to the Milky Way, that would be recovered by
an ideal, volume limited survey. As the available SLSN sample has been
gathered from non-ideal, flux limited surveys, we discuss how we can
fit our volume limited pAMD models to the observed sample after
filtering the models for flux limited selection bias. There could
always be additional bias, particularly the unknown publication bias,
that we cannot account for. Given the small ROTSE-IIIb sample size,
however, the precision of rates may yet be dominated by statistical
errors, and if our pAMD models prove deficient in lower luminosity
events, our rates will still stand as lower limits on the larger
population.

\subsection{SLSN Light Curves}\label{lc}

We define a light curve sample using published photometry of SLSNe.
The sample consists mostly of the 18 events listed in
\citet{galyam2012}. The light curve for SN\,1999as \citep{knop1999}
has not been published and we are thus unable to include it. The light
curve sample includes the ROTSE-IIIb discoveries above. We have added
unpublished observations of SN\,2006tf, Dougie, and another ROTSE-IIIb
discovery, SN\,2010kd \citep{vinko2010}, a SLSN-I that was discovered
after the survey period considered here. Table\,\ref{table:lcsample}
lists the SLSNe considered.

\begin{table*}
\centering
\caption{Light curve sample of SLSN-like events.}
\label{table:lcsample}
\begin{tabular}{lllcrrl}
\hline
Name & Type & Abs. Mag$^{a}$ & $\Delta m_{40}$ & Effective & Effective & Reference \\
     &    &    &   & Vol.-Time$^{b}$ & Vol.-Time$^{c}$ & \\
 \hline
    SN2003ma &  SLSN-II & $-21.58$ & 0.55 &  0.0296 &  0.0286 & \citet{rest2011} \\
    SN2005ap &   SLSN-I & $-22.15$ & 1.36$^{d}$ &  0.0503 &  0.0288 & \citet{quimby2007c} \\
     SCP06F6 &   SLSN-I & $-22.11$ & 1.41 &  0.0561 &  0.0341 & \citet{barbary2009} \\
    SN2006gy &  SLSN-II & $-20.75$ & 0.64 &  0.0113 &  0.0349 & \citet{smith2007} \\
    SN2006oz &   SLSN-I & $-21.67^{d}$ & ... &     ... &     ... & \citet{leloudas2012} \\
    SN2006tf &  SLSN-II & $-20.53$ & 0.15 &  0.0103 &  0.0432 & \citet{smith2008} \\
    SN2007bi &   SLSN-I & $-21.02$ & 0.35 &  0.0182 &  0.0450 & \citet{galyam2009} \\
    SN2008am &  SLSN-II & $-21.77$ & 0.24 &  0.0653 &  0.0503 & \citet{chatzopoulos2011} \\
    SN2008es &  SLSN-II & $-22.02$ & 0.87 &  0.0510 &  0.0269 & \citet{gezari2009,miller2009} \\
    SN2008fz &  SLSN-II & $-21.91$ & 0.70 &  0.0494 &  0.0311 & \citet{drake2010} \\
      Dougie &  SLSN-I? & $-22.50$ & 2.66 &  0.0332 &  0.0129 & Vinko et al. (in prep.) \\
    PTF09atu &   SLSN-I & $-21.59$ & 0.52 &  0.0385 &  0.0458 & \citet{quimby2011} \\
    SN2009jh &   SLSN-I & $-21.68$ & 0.69 &  0.0397 &  0.0416 & \citet{quimby2011} \\
    PTF09cnd &   SLSN-I & $-21.90$ & 0.55 &  0.0565 &  0.0452 & \citet{quimby2011} \\
   CSS100217 & SLSN-II? & $-22.79$ & 0.18 &  0.2870 &  0.0546 & \citet{drake2011} \\
    SN2010gx &   SLSN-I & $-21.54$ & 1.54 &  0.0220 &  0.0288 & \citet{pastorello2010,quimby2011} \\
    PS1-10ky &   SLSN-I & $-21.92$ & 1.15 &  0.0433 &  0.0317 & \citet{chomiuk2011} \\
   PS1-10awh &   SLSN-I & $-21.90$ & ... &  0.0356 &  0.0273 & \citet{chomiuk2011} \\
    SN2010kd &   SLSN-I & $-21.08$ & 0.28 &  0.0245 &  0.0562 & Vinko et al. (in prep.) \\
\hline
\end{tabular}

$^{a}$Pseudo-absolute magnitudes are in the unfiltered ROTSE-IIIb rest
frame system and not corrected for host absorption.

$^{b}$ Effective volume-time in units of Gpc$^{3}$\,yr\,$h_{71}^{-3}$
using only the object's light curve and an assumed intrinsic Gaussian
distribution of peak magnitudes ($\sigma=0.3$\,mag ).

$^{c}$ Effective volume-time in units of Gpc$^{3}$\,yr\,$h_{71}^{-3}$
using only the object's light curve scaled to the average peak
magnitude of the group ($-21.7 \pm 0.4$\,mag for SLSN-I and $-21.4 \pm
0.6$\, mag for the SLSN-II).

$^{d}$Estimate based on extrapolation of the light curve.

\end{table*}

We have gathered the observed photometry in each of the available pass
bands, and we use these to estimate the rest frame absolute magnitudes
in the ROTSE-IIIb band pass. When multiple bands are available, we
select the subset of (sometimes several) pass bands that best overlap
with the rest frame ROTSE-IIIb response. We convert the observed
magnitudes to pseudo-absolute magnitudes with $M = m_X - \mu - A_X -
k_X$, where $m_x$ is the observed magnitude in band $X$, $\mu$ is the
distance modulus, $A_X$ is the Galactic extinction in band $X$, and
$k_X$ is the k-correction term for converting the observer frame pass
band into the rest frame ROTSE-IIIb band. We do not include a term to
correct for host absorption, hence our values are ``pseudo-absolute''.
Our unfiltered ROTSE-IIIb magnitudes are calibrated against the
USNO-B1.0 R2 magnitudes, and we include the conversion from observed
magnitudes in the AB system to the Vega system in the k-correction
term when applicable.

To properly estimate the k-corrections, we need to know the intrinsic
spectral energy distribution (SED) of each target as a function of
time. As this information is not always available, we make the
following approximations.

\begin{figure}
\includegraphics[width=\linewidth]{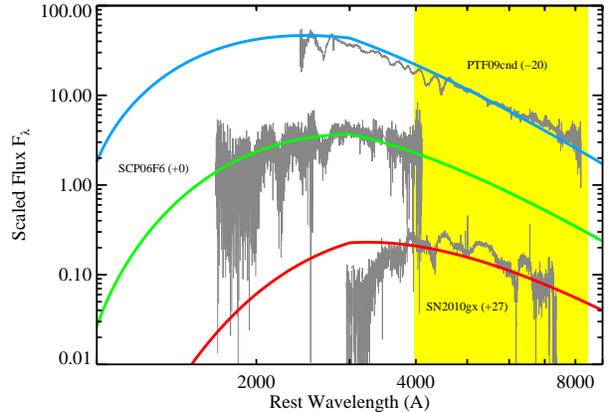}
\caption{ SLSN-I spectral templates (thick blue, green, and red lines)
  compared to the actual spectra of supernovae at 3 different phases
  (20 days before maximum, near maximum, and 27 days after maximum in
  the rest frame ROTSE-IIIb system). The spectra of PTF09cnd,
  SCP\,06F6, and SN\,2010gx are taken from \citet{quimby2011},
  \citet{barbary2009}, and \citet{pastorello2010}, respectively. The
  yellow shaded area marks the ROTSE-IIIb rest frame band pass.  }
\label{fig:slsn1-sed}
\end{figure}

For the SLSNe-I, we assume an augmented, cooling black body for the
SED. Using the analysis of \citet{chomiuk2011}, we fit a cooling curve
to the sample of SLSN-I with spectroscopically derived black body
temperatures. We add the best fit black body temperature measured from
the spectra of SN\,2007bi about 48 rest frame days after peak, which
fits in with the overall cooling trend. To approximate the effects of
metal absorption in the UV \citep{pastorello2010,chomiuk2011}, we
scale the template flux below 3000\,\AA\ by a linear function that
grows from 0 at 912\,\AA\ to 1 at 3000\,\AA. We compare the observed
spectra of several SLSN-I covering a range of rest frame wavelengths
to our assumed templates at similar phases in Figure
\ref{fig:slsn1-sed}. As we are ultimately interested in the response
over ROTSE-IIIb's very broad, unfiltered band pass (or the combination
of multiple broad bands), we assume that the effects of individual
line features missing from the templates to be minor. For example, if
we compare to the k-corrections derived from the actual spectrum of
PTF09cnd shown in Figure \ref{fig:slsn1-sed} to those derived with our
SED model, then, assuming a redshift of $z=0.17$, our SED model (at
the appropriate phase) gives k-corrections in the $g$ band that are
$\sim 0.1$\,mag too faint, accurate to less than 0.01 mag in the $r$
band, and less than 0.1 mag too bright in the $i$ band. Averaging the
three bands should thus result in reasonably accurate absolute
magnitudes (we will adopt a 0.2\,mag offset to estimate systematic
error in our rates).

Near maximum light, which is the most important phase for our rate
calculation, the features present in the rest frame ROTSE-IIIb are
very weak in comparison to normal supernovae, so the temperature of
the black body is the main concern. The results of \citet{chomiuk2011}
suggest that SLSN-I follow similar cooling curves, although there may
be a range of peak temperatures. We note that the SDSS-II photometry
of SN\,2006oz indicates a nearly constant photospheric temperature with
time \citep{leloudas2012}, and the earliest phases are cooler than our
simple model. The peak of this particular event is not well
constrained, so we cannot include it in our efficiency calculations.
 
\begin{figure}
\includegraphics[width=\linewidth]{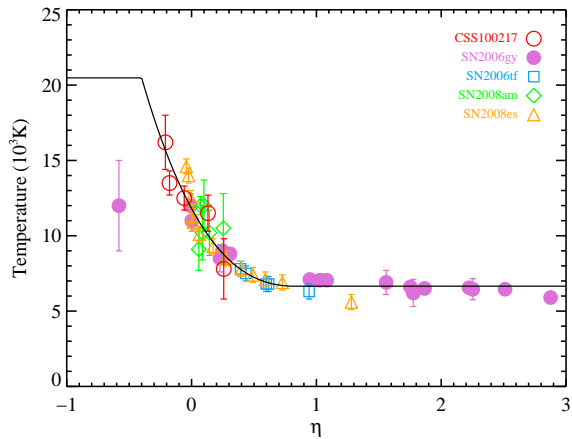}
\caption{ Blackbody temperature of SLSNe-II as a function of the
  parameter, $\eta$, described in the text. The symbols with error
  bars plot individual measurements from \citet{drake2011},
  \citet{smith2007}, \citet{smith2008}, \citet{chatzopoulos2011}, and
  \citet{miller2009}. The black line is the best fit cubic polynomial
  for $-0.4 < \eta < 0.8$ and a constant above and below these
  limits. }
\label{fig:slsn2-bbtemps}
\end{figure}

For the SLSNe-II, we assume a simple Planck function for the SED. This
is motivated by events like SN\,2008es, which was well observed and
found to be well fit by a black body \citep{miller2009,gezari2009},
and other SLSNe-II like SN\,2006tf \citep{smith2008} and SN\,2008am
\citep{chatzopoulos2011}, which, apart from their hydrogen emission
lines, are reasonably well approximated by black bodies. The
temperature measured at peak for SN\,2008es ($\sim$14,000\,K) is much
higher than for SN\,2006tf ($\sim$8000\,K), but both objects cool in a
similar manner as they fade. We find that the temperature evolution of
SLSNe-II can be parameterized as $T(\eta) = 11.8 - 15.3\eta +
14.4\eta^2 - 4.2\eta^3$ (in units of $10^3$\,K; see
Fig. \ref{fig:slsn2-bbtemps}). The parameter, $\eta$, is a function of
the peak absolute magnitude in the rest frame ROTSE-IIIb band,
$M_{peak}$, and the decline from peak, $\Delta M$, such that $\eta =
0.25 (M_{peak} + 22) - \Delta M$ before peak and $\eta = 0.25
(M_{peak} + 22) + \Delta M$ after. The $\eta$ parameter thus
corresponds to the drop in magnitudes below peak (with negative values
for pre-maximum phases) normalized to the behavior of a $M=-22$
source. We limit $\eta$ to $-0.4 < \eta < 0.8$, so the temperatures
can be up to 20000\,K early on, but never below 6700\,K at later
times. We have also experimented with other {\it ad hoc}
parameterizations of the temperature evolution, but these made no
significant changes to our results.

\begin{figure}
\includegraphics[width=\linewidth]{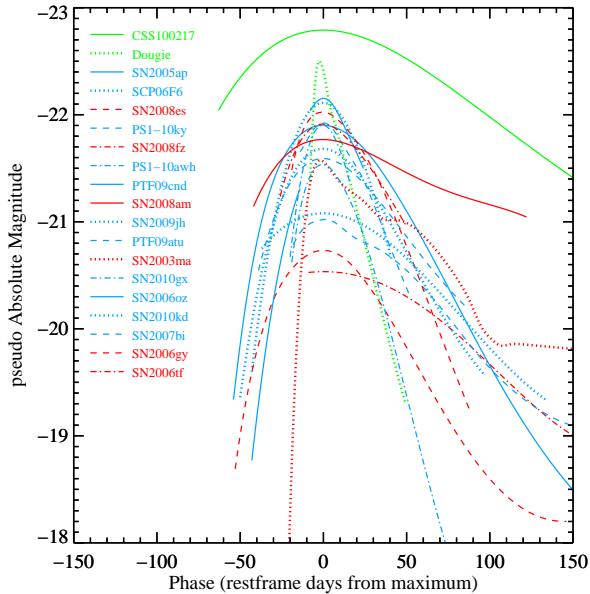}
\caption{ Light curves of SLSN-like events in the rest frame unfiltered
  ROTSE-IIIb band pass. Blue curves are for SLSN-I, and red curves
  mark SLSN-II. The green curves are for two objects of uncertain
  origin. }
\label{fig:lcs}
\end{figure}

To construct the final light curve templates, we fit low order
polynomials to the photometry (after conversion to the rest frame
ROTSE-IIIb band pass as described above). In some cases, we smoothly
combine fits to the rising and declining phases to better capture
asymmetries in the light curves without resorting to higher order
polynomials. Since the date of maximum light depends on the
k-correction, and the k-corrections depend on the phase relative to
maximum light (and also the peak magnitude in the case of SLSNe-II),
we begin with an initial guess for the date of maximum and iterate the
k-corrections and polynomial fits until they converge. The resulting
templates are shown in Figure~\ref{fig:lcs}. Note that our light curve
templates are constrained by observations taken, on average, every 3.5
rest frame days within 20 days of maximum light (without double
counting the multiple bands usually obtained on the same night). 

Several of the supernovae in our light curve sample have both
unfiltered ROTSE-IIIb photometry and filtered observations. After
k-corrections the ROTSE-IIIb measurements are about 0.1\,mag fainter
than the (similarly corrected) filtered photometry of SN\,2006gy and
SN\,2008es, and the ROTSE-IIIb measurements are about 0.1\,mag
brighter for Dougie and SN\,2010kd. The largest discrepancy seen is
for SN\,2006tf, for which the ROTSE-IIIb measurements are about
0.25\,mag brighter than the filtered photometry available from
\citet{smith2008} after k-corrections. The observed ROTSE-IIIb
magnitudes agree with the observed R-band measurements to within
0.1\,mag over the same range (out to about 100 days after maximum
light), so the discrepancy in k-corrected, absolute magnitudes is
probably due in part to the k-corrections themselves (as noted above,
we will assume a 0.2\,mag systematic error in magnitudes for the rate
measurements). In the case of SN\,2006tf, we choose to shift our
unfiltered photometry to match the filtered measurements. This
preserves the constraints on maximum light from the ROTSE-IIIb
observations, which precede the filtered data by up to 1 month.

CSS100217 \citep{drake2011} appears to be distinctly brighter than the
rest of the objects considered, and, integrating its light curve, it
radiated far more energy than the others over the period shown in
Figure~\ref{fig:lcs}. It was also located precisely coincident with a
known AGN (SDSS\,J102912.58+404219.7; cf.
\citealt[][]{greene2007}). \citet{drake2011} favor classifying this
object as a SLSN-II because its variability is larger than normal AGN
flares, but given that its amplitude is also larger than even the
extreme supernovae considered here, a similar argument could be used
to disfavor classification as a supernova (see further discussion in
\S\ref{conclusions}). Similarly, Dougie is brighter than the other
remaining events, it rises and fades faster, and, as noted above, it
shows a spectral evolution that is distinct from most SLSNe. While
each of these may yet prove to be supernovae, the evidence against
this interpretation is considerable. We will thus calculate the SLSN
rates below with and without these questionable events.

\subsection{SLSN Pseudo-Absolute Magnitude Distribution}\label{pamd}

From the light curve templates constructed above, we derive the
approximate pseudo-absolute magnitude distribution of SLSNe-I and
SLSNe-II. Ideally, one would like to have a volume limited sample from
which to derive these distributions. This is simply not available, and
we are further limited by the small numbers in the flux limited
samples. Given these limitations, it may be tempting to simply take
the peak magnitudes from the light curve sample, measure the average
and standard deviation, and then assume a Gaussian distribution with
these parameters (and we do consider such distributions for the rate
calculations below for reference). However, this will generally be
problematic since the observed sample is flux limited and thus lacking
representation from the lower luminosity members of the true
population. The mean and standard deviation from the observed samples
will thus be biased. Additionally, real, astrophysical sources that
lie in star-forming galaxies will be obscured by dust. This will skew
an intrinsically Gaussian distribution and produce an extended faint
tail to the pAMD.

We thus attempt to derive realistic pAMDs from the data
available by assuming the intrinsic populations obey a Gaussian
distribution for simplicity and including the skewing effects of
dust. To translate these ideal models into the distributions recovered
by a flux limited search, we simply weight the ideal distributions by
the relative volumes expected from such a survey. We can then find the
best pAMD (the one that would be found from an ideal, volume limited
survey) by varying the input model until the best match is found
between the weighted model and the observed sample. As the model
separates the effects of dust from the intrinsic scatter, insights
into the later may be gained if the correct dust model is
used. SN\,2006oz was not observed at maximum light, so we must exclude
it from what follows.

The sample is based on discoveries drawn from several different
surveys, each with their own selection biases. The ultimate selection
function for our light curve sample is thus complex and largely
unknown. To simplify matters, we assume the aggregate population is
selected from a hypothetical, flux limited survey. This should be a
reasonable assumption since the contributing surveys are flux limited
(not targeting specific luminosity ranges or host galaxy types) at
their core, with spectroscopic follow-up adding a second roughly flux
limited layer to the discovery process. There could be additional
biases, however, such as a preference for publication of the higher
luminosity events, that may further skew our light curve sample. Such
effects should typically limit the inclusion of lower-luminosity
events (in particular, we note that the sample from
\citealt{galyam2012} is based on events that peak above $-21$\,mag in
any given photometric system). This could lead us to overestimate our
survey efficiency and thus underestimate the rates.

We consider simple models for the pAMDs based on intrinsic luminosity
functions that follow a Gaussian distribution combined with host
absorption drawn from an exponential distribution of the form $P(A_V)
\propto e^{-A_V/\tau}$ (the V-band is the best match to the ROTSE-IIIb
band when redshifted to $z\sim0.2$). We adopt $\tau = 0.6$ for the
SLSN-I and SLSN-II populations because these events are very likely to
be connected with massive stars and the absorption distribution
expected for objects originating from such stars is approximately fit
with this exponential distribution \citep{hatano1998}. As shown below,
this choice also provides a reasonable match to the observed
distributions of SLSN in the light curve sample after transforming the
pAMD models to the distributions expected from a flux limited survey.

\begin{figure}
\includegraphics[width=\linewidth]{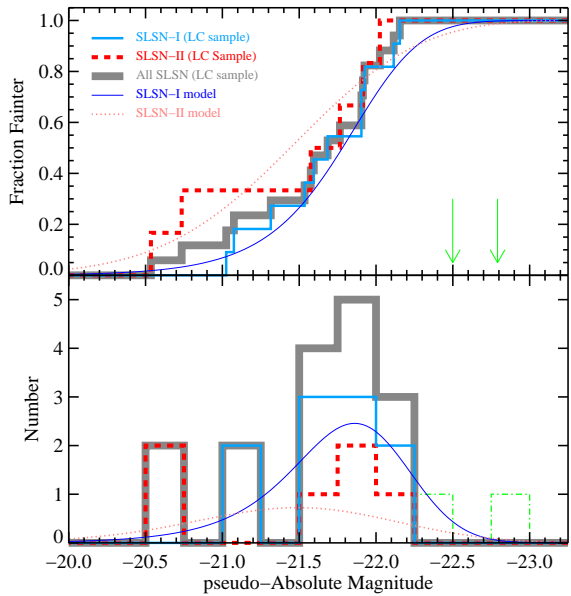}
\caption{ Peak pseudo-absolute magnitudes for SLSN-I (blue) and
  SLSN-II (dashed red) in the light curve sample. SN\,2006oz is not
  included because its peak magnitude is not well constrained. The
  locations of CSS100217 and Dougie, which may or may not actually be
  SLSNe, are marked with the green arrows on the top plot (cumulative
  distribution) and dashed-dotted green boxes on the lower plot
  (binned distribution). The dark blue and dotted pink curves show the
  models used in the survey efficiency calculations adjusted for the
  selection effects of a flux limited survey for comparison to the
  observed sample. The SLSN-I model has an intrinsic Gaussian width of
  just $\sigma=0.3$\,mag. The skewed faint end tail of the model is a
  result of the assumed host absorption distribution.  }
\label{fig:pamd}
\end{figure}

With $\tau$ fixed, we vary the model's peak (intrinsic) magnitude,
$M$, and its Gaussian width, $\sigma$ to find the best match to the
light curve sample. We measure the maximum offsets in the cumulative
distributions of the models and data and we record the
Kolmogorov-Smirnov probability that the observed distribution was
drawn from each model. Given the limited SLSN-I and SLSN-II sample
sizes, a variety of values are plausible, but we find the best
agreement for the SLSN-I sample around $M=-22.0$ and $\sigma=0.3$,
while the SLSN-II population is best matched with values around
$M=-21.4$ and $\sigma=0.6$. The peaks of the intrinsic distributions
cannot be much brighter (e.g. $M \simgt -22.3$ for the SLSN-I), but
fainter values are plausible if the widths of the distributions are
larger. We did not allow the absorption distribution to vary, but
smaller values of $\tau$ would decrease the expected number of the
lower luminosity events, which is less preferred (but not
excluded). We considered observed distributions both with and without
CSS100217 and Dougie, but the differences are negligible. The
resulting pAMD models are shown in Figure~\ref{fig:pamd}.

We have included SN\,2007bi and SN\,2010kd in the SLSN-I light curve
sample even though their light curves decline (and likely rise) much
more slowly than the other SLSN-I, which could point to a different
physical origin (\citealt{galyam2009} have concluded that SN\,2007bi
is a radioactively powered, pair-creation supernova, but this
possibility is excluded for others in the SLSN-I sample). Even with
these events, the width of the intrinsic luminosity function could be
rather narrow. If there is a publication bias against lower luminosity
SLSN-I, then the true distribution could be broader than the available
evidence suggests. It is also possible to explain the observed scatter
in SLSN-I peak magnitudes from the effects of dust alone, with zero
intrinsic dispersion (assuming the dust model adopted above). 

We estimate the uncertainty in the peaks of our distributions by
varying the temperatures assumed in calculating the k-corrections for
our light curve sample. To determine the systematic error, we adjust
the model temperatures by $\pm1000$\,K, and add to this an estimate of
the statistical uncertainty in the peak magnitudes. 

\section{The SLSN Rates}\label{rates}

In this section, we combine the ROTSE-IIIb sample presented in section
\ref{sample} and the light curves and pAMDs infered from the published
sample of SLSNe (presented in section \ref{lcsample}) to derive the
volumetric rates of SLSN-I, SLSN-II, and all SLSN-like objects (which
may include additional, rare phenomena that may or may not be true
supernovae). The rates are calculated from,
\begin{equation}\label{eqn:rate}
 R = {N \over \sum_{i}\epsilon_i  V_i t_i}
\end{equation}
\noindent with $N$ the number of events detected, $V_i$ the co-moving
volume element for the $i^{\rm th}$ distance bin, $\epsilon_i$ the
corresponding survey efficiency for that bin, and $t_i$ the proper
time for the survey in each bin. We refer to the denominator in
eqn. \ref{eqn:rate} as the ``effective volume-time'' in what
follows. We first describe the Monte Carlo simulations employed to
calculate the survey efficiency before presenting the actual rates.

\subsection{Monte Carlo Simulations}

Following the procedure described in \citet{quimby2012}, we have run
Monte Carlo simulations to determine how efficient our ROTSE-IIIb
search has been in selecting SLSNe. We define our survey efficiency,
$\epsilon_i$, as the fraction of SLSNe in a given volume that we are
likely to select. There are two main factors in selecting SLSNe: 1)
are the observations deep enough to detect a given SLSNe at a given
distance, and 2) did we observe the appropriate sky location during
the right time to catch the transient during outburst (relevant only
if the observations are deep enough). The first factor results from
the flux limited (i.e. not volume limited) nature of our survey. At a
given distance we may detect some of the more luminous SLSNe but miss
some of the fainter events. The second factor is a result of our
sampling function (or ``cadence''). Due to weather, observing seasons,
and time constraints, we have not observed the same fields every day,
and some fields have far less coverage that others (see figures 1 and
2 from \citealt{quimby2012}), so we could have missed even bright
SLSNe in coverage gaps. We thus perform a Monte Carlo simulation in
which we compare SLSNe light curves at various sky locations,
distances, and random explosion dates to the actual survey data to
determine what fraction of simulated events we are likely to have
selected.

We determine our selection efficiency in distance bins from 40 to
4000\,Mpc (beyond this our selection probability is zero, as we will
show below). For each distance bin, we convert the absolute magnitude
light curves from section \ref{lcsample} to their expected (or
``observed'') magnitudes by adding the distance modulus, the same
crude k-corrections described above, and the Galactic extinction from
\citet{sfd1998} for a given sky position. Next we randomly select a
date for maximum light between Nov 1, 2004 and January 31, 2009. The
observed magnitude can then be computed for any day over the survey
period. We compare the expected magnitudes for our simulated SLSNe to
the actual data recorded to determine the fraction of events likely to
have been selected. To do this, we use the detection efficiency curves
from \citet{quimby2012}, which give the probability of detecting an
object as a function of magnitude. These detection efficiency curves
take into account the seeing quality for the image as well as the
``limiting magnitude.''

In the actual survey, the selection of targets was done by humans
reviewing an automatically generated ``short list'' of candidates that
was constructed from targets with at least a $5\sigma$ detection in a
nightly image stack, and at least $2.5\sigma$ detections in the first
and second halves of the night (in addition to other cuts on shape
parameters and increase in brightness; see \citealt{quimby2012} for
full details). Thus, in our simulations we calculate the detection
probabilities on each of the three stacks, draw three random numbers
(uniformly distributed between 0 and 1), and we only count a simulated
source as recovered if each of the three probabilities is larger than
their respective random draws.

As in \citet{quimby2012}, we compare the expected distributions in the
distance, observed peak magnitudes, pseudo absolute magnitudes, and
the number of nights a simulated source is detected from the
simulations to the observed data. Given our small sample size, we lack
the statistical power to identify differences between how the
simulated sources are selected versus the actual survey data; however,
we do note that not one event in our observed sample was detected on
fewer than 5 nights. In contrast, the simulations typically predict
that $\sim$60\% of our discoveries should only be detectable on 4 or
fewer nights. \citet{quimby2012} found a similar disagreement between
the models and data for a larger sample of Type\,Ia supernovae, which
could be resolved by considering only the events (real or simulated)
detected on 5 or more nights. The dearth of objects detected on just
1-4 nights is presumably attributed to human bias in the selection of
targets. We assume a similar effect for our SLSN sample and limit the
models and data to events detected on 5 or more nights. With this cut,
the four model distributions considered appear to be well matched with
the observations. If we were to ignore this selection cut, the rates
found below would be lowered by a factor of 2.5.

\subsection{SLSN-I Rates}

We have only one discovery, SN\,2005ap, from which to derive the
SLSN-I rate. Obviously, our rate will be subject to the eccentric
properties of small number statistics, and care should be taken in
interpreting these results.

We first perform Monte Carlo simulations using the SLSN-I light curves
from section \ref{lcsample} (excluding Dougie for the moment) and the
the SLSN-I pAMD model derived in section \ref{pamd} to determine our
effective volume-time. Under these assumptions, the effective
volume-time of the ROTSE-IIIb search is
$0.0315$\,Gpc$^{3}$\,yr$^{1}$\,$h_{71}^{-3}$, which gives a rate of
$(32^{+77}_{-26})$\,events\,Gpc$^{-3}$\,yr$^{-1}$\,$h_{71}^{3}$, where
the error bars account for only the statistical Poisson fluctuations
\citep{gehrels1986}.

To estimate our systematic error, we recalculate our survey
efficiencies with the input pAMDs shifted 0.2\,mag brighter or
fainter. This is to account for possible systematic error in our
k-corrections, which may bias the peak magnitude distributions. Given
our small sample size, the statistical errors dominate the
systematics, which we find to be
$^{+10}_{-7}$\,events\,Gpc$^{-3}$\,yr$^{-1}$\,$h_{71}^{3}$ for the
SLSN-I. We note, however, that due to selection (and particularly
publication) bias, our pAMD may not accurately represent the full
population of events physically connected to the SLSNe-I. In this
case, our rates are valid only for the luminosity range considered and
are simply lower limits for the greater population.

\begin{figure}
\includegraphics[width=\linewidth]{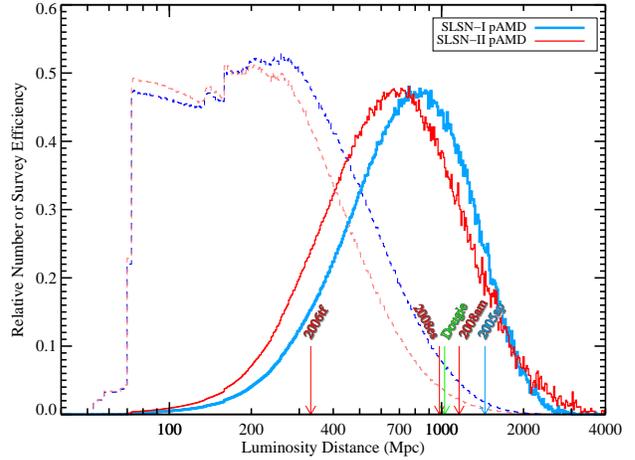}
\caption{ Survey efficiencies (dashed lines) and the relative distance
  distributions (solid lines) of SLSN-I (blue curves) and SLSN-II (red
  curves) expected from these. The luminosity distances for the SLSNe
  in the ROTSE-IIIb sample are indicated with arrows. }
\label{fig:efficiency}
\end{figure}

Figure~\ref{fig:efficiency} shows our survey efficiency as a function
of luminosity distance. The expected distance distribution of SLSN-I
is also shown with the solid blue curve. This latter curve is
calculated by multiplying the efficiency curve by the relative volume
and proper time in each distance bin. Although the sensitivity curve
drops off above $\sim$300\,Mpc, this is initially compensated for by
the nearly cubic rise in volume with distance. Due to this rise in
volume, we are sensitive to SLSN-I out to more than 2000\,Mpc even
though we may detect less than 1\% of this population. We expect to
find about 97\% of our SLSN-I between 200 and 2000\,Mpc (roughly $0.05
< z < 0.4$), and half should be farther than about 800\,Mpc. Note that
at $D_L \sim 1440$\,Mpc, SN\,2005ap falls comfortably with in our
expectations. The average effective redshift for our SLSN-I rate
measurement is thus $z=0.17$.

As a check, we also provide effective volume-times assuming only the
light curves of individual supernovae and two separate assumptions on
the peak magnitude distribution in table \ref{table:lcsample}. The
values in the fifth column assume that the distribution is a simple
Gaussian (no skewing by host absorption) that takes the reported peak
magnitude of the supernova for the peak of the distribution and
assumes a width of $\sigma=0.3$\,mag. This width was chosen to
represent the (assumed) intrinsic scatter in events otherwise
identical to each object listed and is based on the intrinsic scatter
of SLSN-I estimated above. For SN\,2005ap, the effective volume-time
is $0.0504$\,Gpc$^{3}$\,yr$^{1}$\,$h_{71}^{-3}$ under these
assumptions. From equation \ref{eqn:rate}, this leads to a rate of
$(20^{+48}_{-16})$\,events\,Gpc$^{-3}$\,yr$^{-1}$\,$h_{71}^{3}$
(statistical error only).

The sixth column in table \ref{table:lcsample} gives the effective
volume-times again assuming only the single light curve of the
supernova and a simple, uncorrected Gaussian distribution but with the
peak and width set by the average and standard deviation of the
apparent SLSN-I sample, respectively ($M = -21.7$ and $\sigma =
0.4$\,mag). With these assumptions, the effective volume-time drops to
$0.0288$\,Gpc$^{3}$\,yr$^{1}$\,$h_{71}^{-3}$. The rate for this
distribution is then
$(35^{+84}_{-29})$\,events\,Gpc$^{-3}$\,yr$^{-1}$\,$h_{71}^{3}$.

Our original measurement is fully consistent with these two later
assessments. Again, the small sample size (one object) results in large
statistical errors which dominate systematic concerns. As the
procedure used to estimate the pAMD model is generally less prone to
bias than the later, simple Gaussian assumptions, we retain its
effective volume-time for our best measurement of the SLSN-I rate.

If we count Dougie as a SLSN-I, then the effective volume-time
(including Dougie and all SLSN-I light curves in the Monte Carlo
simulations) decreases slightly to
0.0291\,Gpc$^{3}$\,yr$^{1}$\,$h_{71}^{-3}$. The combined rate for
these two events is then
$(68^{+94}_{-44})$\,events\,Gpc$^{-3}$\,yr$^{-1}$\,$h_{71}^{3}$
(statistical error only).

\subsection{SLSN-II Rates}

Our ROTSE-IIIb sample includes three events with clear signatures of
hydrogen in their spectra: SNe 2006tf, 2008am, and 2008es. Although
they share some similar properties, there are differences as well. In
particular, the available observations of SN\,2008es do not show the
obvious narrow emission lines from hydrogen. These are present in the
spectra of SNe 2006tf and 2008am and suggest an interaction of the
supernova ejecta with slow moving circum-stellar material as a
contributing power source. It could be, however, that SN\,2008es
experienced a similar interaction prior to the first available
spectra \citep[cf.][]{moriya_tominaga2012}.

Given the apparent differences, one could follow a similar procedure
as outlined above and calculate three separate rates using the
effective volume-times in Table \ref{table:lcsample}. Here we use the
full ensemble of SLSN-II light curves available in the light curve
sample and the pAMD model from section \ref{pamd} to place limits on
the bulk, hydrogen rich SLSN population.

Using the SLSN-II pAMD model, we find an effective volume-time of
$0.0199$\,Gpc$^{3}$\,yr$^{1}$\,$h_{71}^{-3}$ (this would be 6\% larger
if the light curve of CSS100217 were included). From equation
\ref{eqn:rate}, this gives a SLSN-II rate of
$(151^{+151}_{-82})$\,events\,Gpc$^{-3}$\,yr$^{-1}$\,$h_{71}^{3}$
(statistical error only) for our sample of three events. Performing
the same check described above, we estimate a systematic error of
$^{+52}_{-33}$\,events\,Gpc$^{-3}$\,yr$^{-1}$\,$h_{71}^{3}$. Again, we
stress that there could be lower luminosity events that are physically
related to those considered in determining the pAMD. In this case, our
rate is a lower limit on the greater population.

From Figure \ref{fig:efficiency}, it is apparent that like the SLSN-I,
we are sensitive to SLSN-II in roughly the 200 to 2000\,Mpc range. The
average redshift for our SLSN-II rate measurement is $z=0.15$.

\subsection{SLSN-Like Rates}

Finally, we measure the total rate of SLSN-like events including
possible tidal disruption events or rare, supernova-like outbursts
from (uncatalogued) AGN. Our survey was not intentionally biased
against events located near galaxy nuclei (for example, SN\,2006gy was
found less than 0.3 pixels from the core if its host), but we did
exclude a few sources detected coincident with known AGN. Our
detection efficiency was also calculated for blank sky locations only,
so it may be an overestimation for sources on top of bright galactic
nuclei. So once again, our rate may be considered a lower limit.

We include Dougie with the other 4 background events and assume the
effective volume-time is simply given by the average of the SLSN-I and
SLSN-II values found above (including Dougie and CSS100217 in the
Monte Carlo simulations). Since the pAMDs for these two groups are
broadly similar, this should be a reasonable approximation. With these
assumptions, the total SLSN-like rate is
$(199^{+137,+65}_{-86,-41})$\,events\,Gpc$^{-3}$\,yr$^{-1}$\,$h_{71}^{3}$,
where the first and second errors quoted are statistical and
systematic, respectively.

\section{Conclusions and Discussion}\label{conclusions}

We have produced a set of SLSN light curve templates corrected to the
rest frame, unfiltered ROTSE-IIIb band pass and used these to estimate
the pseudo-absolute magnitude distributions of SLSN-I and
SLSN-II. With these and our sample of ROTSE-IIIb discovered SLSN (and
a SLSN-like event), we have calculated the volumetric rates in the
local volume ($\overline{z} \sim 0.2$). Compared to the $z \sim 0.2$
core collapse supernova (CCSN) rate \citep[cf.][]{botticella2008}), we
find that there is about one SLSN-I or SLSN-II event for every 400
to 1300 CCSNe, or one SLSN-I for every $1000$ to $2 \times 10^4$
CCSNe (these fractions may be halved if the full population of faint
or obscured CCSNe is included; \citealt{horiuchi2011}). The total
SLSN-like rate is also similar to the local rate of sub-energetic,
long-duration gamma-ray bursts (LGRB; \citealt{soderberg2006}),
although the SLSN-like and LGRB samples are currently too small for a
definitive comparison.

If the SLSNe rates simply track the cosmic star formation rate (CSFR),
then, using the parameterization of \citet{yuksel2008}, the CSFR and
thus SLSN rate should be about 5 times larger at $z\sim3$ than we have
measured for the nearby ROTSE-IIIb sample. Given the large errors,
this could be roughly consistent with the rate of SLSNe at $z \sim 2$
and 4 \citep{cooke2012}. However, the search performed by
\citet{cooke2012} was likely biased to longer duration events, and
they suggest their discoveries represent only the radioactively
powered subset of SLSNe (and only a fraction of those). In this case,
their high redshift discoveries could imply an enhancement over a
simple scaling by the CSFR when compared to an appropriate fraction of
our low redshift rates, as they have noted.

The SLSN-I in the light curve sample are particularly
interesting. Although these events were observed in a variety of (rest
frame) band passes and we apply only a simplistic k-correction, they
still appear to constitute a surprisingly uniform class when
transformed to the ROTSE-IIIb system (see Fig. \ref{fig:lcs}). The
light curves show some quantitative differences but tend to rise
smoothly over around 50 rest frame days, which is considerably longer
than typical Type\,Ia supernovae. Several SLSN-I also show smooth
declines that almost mirror their pre-maximum behavior. The peak
magnitudes of the SLSN-I in the light curve sample have an average of
$-21.7$\,mag and a standard deviation of $\pm 0.4$\,mag. Our best
estimate of the intrinsic scatter (after accounting for absorption in
the host galaxies and selection bias) is about $\pm 0.3$\,mag, but
this is not well constrained. These values are based on a small sample
(10 events) and change negligibly if Dougie is added. Given that the
peak magnitudes of the SLSN-I sample are relatively tightly
distributed, it is worth considering if this is a mere selection bias
or indicative of a population that may prove useful as standard
candles.

First, we consider the high luminosity cutoff to the distribution.
The published sample of SLSN-I shows a cap in peak absolute magnitudes
close to $M \sim -22$. If there are such events with significantly
brighter peaks they must be exceedingly rare or they would dominate
flux limited searches. It is thus reasonable to conclude that most
SLSN-I will peak around $-22$\,mag or fainter.

Next, we consider the lower luminosity limit for the SLSN-I available
in the light curve sample. Formally, we have used an arbitrary cut on
peak luminosities ($M_{\rm peak} < -21$) to select the SLSN-I for our
light curve sample, but we note that we could use other, luminosity
independent criteria to define this sample. In particular, the SLSN-I
in the light curve sample contain the only supernovae we know of that
show OII features in their maximum light spectra. Other supernovae
with more normal peak luminosities may also display such features at
very early times, but these quickly disappear as the ejecta cool and
oxygen recombines (cf. the spectral series of SN\,2008D presented by
\citealt{modjaz2009}). As a practical matter, this definition is
perhaps less useful than a simple luminosity cut as it requires
reasonably high signal to noise spectra taken at the right phase and
with the proper wavelength coverage to catch the relatively weak OII
features in the 4400 to 5800\,\AA\ range. Indeed, The wavelength
coverage and signal to noise of SCP\,06F6 and PS1-10ky preclude a
definitive assessment of such features at maximum light, but based on
the available spectral information (particularly the high temperatures
implied and the similar features at 2100 to 2700\,\AA), we find this
to be likely. There are no spectra of SN\,2007bi near maximum light,
but we suspect the OII features would have been visible at maximum as
well considering later phase spectroscopic matches to the SLSN-I
sample.

There are no known supernovae of normal luminosities that show OII at
maximum light (events with this feature at maximum will be called
SN\,2005ap-like hereafter), and this signature may uniquely identify a
potentially useful population of high luminosity supernovae. Based on
the considerations below, we find it likely that all SN\,2005ap-like
events will be superluminous because the OII lines require high
temperatures and the long expansion of the ejecta up to maximum light
necessitate a large radius. The maximum light spectra of SLSN-I are
reasonably well approximated by a black body (at least in the optical
range where line blanketing is not an issue; \citealt{pastorello2010},
\citealt{chomiuk2011}; see Fig. \ref{fig:slsn1-sed}), so the peak
luminosity of SLSNe-I will be approximately, $L = 4 \pi R^2 \sigma
T^4$ (with a small dilution term to correct for the departure from a
black body, which we neglect).  \citet{hatano1999} have shown that OII
features are temperature sensitive. These lines should be replaced by
the usual OI lines for temperatures below about 12000\,K, and by OIII
lines for temperatures above 15000\,K. The typical expansion velocity
of supernovae, $10^4$\,km\,s$^{-1}$ over a one month period gives a
radius of $\sim 3 \times 10^{15}$\,cm. Thus any supernova that remains
hot enough to display prominent OII features after a one month rise to
maximum light should have a peak absolute magnitude brighter than
about $M=-21$.

More detailed theoretical modeling, beyond the scope of this paper, is
required to verify this simplified picture and determine if all
SN\,2005ap-like supernovae must necessarily reach such high
luminosities. If the observed floor to the peak magnitude distribution
proves to be physical and not just a selection effect, then the peak
magnitudes of SN\,2005ap-like supernovae are confined to a relatively
narrow range. This gives potential for SN\,2005ap-like supernovae to
one day serve as standard candles visible as far as $L_{*}$ galaxies
can be seen and clearly warrants further study.

\begin{figure}
\includegraphics[width=\linewidth]{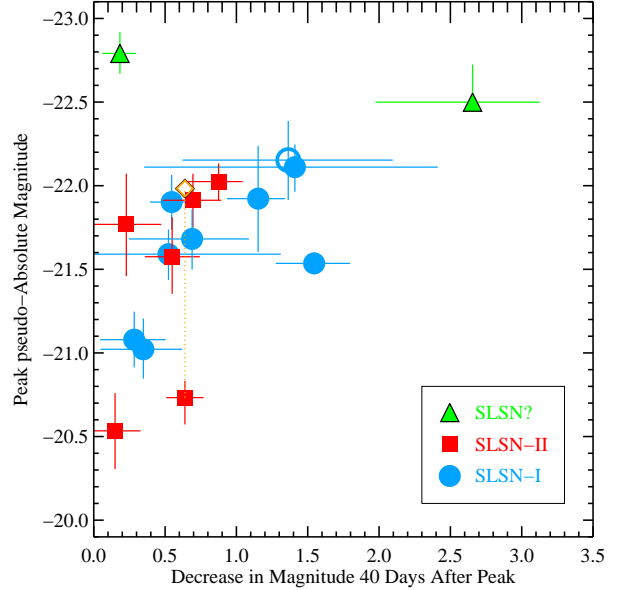}
\caption{ Photometric decline in magnitudes between peak and 40 days
  after peak for the SLSN light curve sample. The open orange diamond
  shows where SN\,2006gy would lie if corrected for host
  absorption. The open circle marks SN\,2005ap, which was not observed
  on day $+40$ or after; the $\Delta m_{40}$ shown for SN\,2005ap is
  an extrapolation of the earlier light curve. Errors include the
  estimated systematics.  }
\label{fig:dm40}
\end{figure}

We next consider if SLSN-I obey a luminosity-decline relation analogous
to the Phillips relation observed for the standard bearer of standard
candles, Type\,Ia supernovae. In Figure~\ref{fig:dm40}, we plot the
change in magnitude between peak and 40 days after peak, $\Delta
m_{40}$, against peak pseudo-absolute magnitude. SLSNe may be better
differentiated in their declines from peak at even later phases, but
the observations are harder to come by. Errors shown are the
estimated statistical error plus an estimate of the systematic error
based on changing the temperature of the SED used in the k-correction,
as discussed above.

Using the Bayesian linear regression routine, {\tt linmix\_err.pro}
\citep{kelly2007}, we do not find a significant correlation for 9
SLSN-I plotted in Figure \ref{fig:dm40}. If we include Dougie,
however, there may be a weak correlation between the peak magnitudes
and decline parameter in the sense that brighter events have faster
declines. This may simply be a selection effect, however, which would
tend to remove the fainter, faster declining events.

Figure \ref{fig:dm40} also includes the SLSN-II from the light curve
sample. It is interesting to note that if the peak magnitude of
SN\,2006gy is corrected for host absorption, then the combined
distribution of SLSN-I and SLSN-II peaks sharply at $M \sim -21.8$
with SN\,2006tf the only SLSN-II falling below\footnote{SN\,2006tf was
  included in the sample of \citet{galyam2012} based on its peak above
  $-21$\,mag as suggested by preliminary ROTSE-IIIb estimates, but our
  revised analysis and dimming to match the filtered photometry places
  it below this cutoff}.  It is unclear if this event is the low
luminosity tail of the SLSNe-II population or the bright tail of the
``normal'' Type\,IIn supernova distribution. Indeed, it is unclear if
two such distinct populations exist in nature, or if there is simply a
publication bias against Type\,IIn supernovae of moderately bright
luminosities. Any such distinction may be borne out by the larger
samples of normal supernovae and SLSNe currently being gathered by
searches such as PTF, Pan-STARRS, the Catalina Real-Time Transient
Survey \citep{drake2009}, and La Silla-QUEST \citep{hadjiyska2012}.

The discovery of such rare events, however, may be hindered by
contamination from other transient phenomena. The ROTSE-IIIb sample
includes about ten times more SNe\,Ia than SLSNe-like events even
though the SLSNe are observable over a much larger volume because the
ratio of the rates is even larger. There is also the worry that other
rare transient phenomena, such as tidal disruption flares or an
outburst from an AGN with a supernova-like light curve, could
contaminate the sample. We potentially have one such source in our
ROTSE-IIIb sample: Dougie. If Dougie proves to be a hydrogen poor
supernova, our SLSN-I rate would roughly double.

The SLSN-II population may similarly be polluted with AGN
activity. \citet{drake2011} favor classifying CSS100217 as a
superluminous supernova even though it radiated far more light than
other SLSN published so far. CSS100217 does show features present in
Type\,IIn supernova, but these are not unique to supernovae and can
also mark an AGN outburst (see \citealt{colgate_cameron1963} and
\citealt{filippenko1989} for a discussion of the similarities between
AGN and SNe\,IIn spectra). Since the host of CSS100217 is known to
harbor an AGN and this is precisely coincident with the transient,
this could simply be a demonstration of rare AGN related
activity. \citet{drake2011} note that amplitude of CSS100217's
outburst stands distinct from typical AGN, but CSS100217 was
specifically selected for this unusual brightening while the
comparison sample was drawn from the SDSS DR3 spectroscopic sample
(i.e. not selected based on variability). The relatively large
effective volume-time for CSS100217-like events listed in Table
\ref{table:lcsample} indicates that such events are either very rare
(least they dominate flux limited surveys, which is not the case), or
they must be selected against due to, for example, the presence of a
coincident AGN. We know of no reason to exclude AGN from having
supernova-like outbursts, so it will be important to measure the rate
of such events to determine their potential contamination to
supernovae samples.

Like CSS100217, SN\,2006gy was located near the core of its host (but
clearly offset in this case). Studies of our own galaxy and M31 have
uncovered a significant population of massive stars located very close
to the central nuclei \citep[][]{genzel2000,lauer2012}. We speculate
that such environments may on occasion harbor dense clouds of gas
accreted from the disk \citep[cf.][]{levin2003}, recent galaxy
mergers, or fed by filaments, and these special environments may yet
give birth to unusual supernovae. These events may be heavily
extincted and thus difficult to detect (e.g. SN\,2006gy suffers from
about 1.25\,mag of absorption in the R-band). This may add to the
confusion with AGN and make the SLSN-II rate difficult to measure.

The rates presented here may be helpful in determining the origin of
SLSNe. For example, using the total star formation rate at z=0.16
\citep{hopkins_beacom2006} and a modified Salpeter IMF
\citep{baldry_glazebrook2003}, we can compare our total SLSN rate to
the expected production rate of stars with $56 M_{\sun} \simlt M_{\rm
  initial} \simlt 84 M_{\sun}$, which is the mass range under which
zero metallicity stars are expected to enter the pulsational pair
instability regime
\citep{chatzopoulos_wheeler2012,yoon_langer2012}. We find that our
SLSN rate could be explained by a small fraction of the stars born in
this mass range (about 1 to 3\% at the $1\sigma$ confidence level). To
put this another way, we can say at the $>10\sigma$ confidence level
that at $z\sim 0.2$, not all stars in this mass range produce
supernovae brighter than $M \sim -21$\,mag; if some of these stars do
produce SLSNe, others must experience alternate fates. These rates
also define a basis for comparison to future, higher redshift studies,
which may help determine if the IMF varies over the distances that
will be probed by forthcoming surveys (i.e. $z\sim4$ with Subaru
Hyper-SuprimeCam).

\section*{Acknowledgments}

This work was supported by Kakenhi Grant-in-Aid for
Young Scientists (B)(24740118) from Japan Society for the Promotion of
Science. Parts of this research were conducted by the Australian
Research Council Centre of Excellence for All-sky Astrophysics
(CAASTRO), through project number CE110001020. ROTSE-III has been
supported by NASA grant NNX-08AV63G and NSF grant PHY-0801007. The
research of JCW is supported in part by NSF grant AST1109801.

\label{lastpage}

\begin{thebibliography}{71}
\expandafter\ifx\csname natexlab\endcsname\relax\def\natexlab#1{#1}\fi

\bibitem[{{Akerlof} {et~al}\mbox{.}(2003){Akerlof}, {Kehoe}, {McKay}, {Rykoff},
  {Smith}, {Casperson}, {McGowan}, {Vestrand}, {Wozniak}, {Wren}, {Ashley},
  {Phillips}, {Marshall}, {Epps}, \& {Schier}}]{akerlof2003}
{Akerlof} C.~W. {et~al.}, 2003, \pasp, 115, 132

\bibitem[{{Baldry} \& {Glazebrook}(2003)}]{baldry_glazebrook2003}
{Baldry} I.~K., {Glazebrook} K., 2003, \apj, 593, 258

\bibitem[{{Barbary} {et~al}\mbox{.}(2009){Barbary}, {Dawson}, {Tokita},
  {Aldering}, {Amanullah}, {Connolly}, {Doi}, {Faccioli}, {Fadeyev},
  {Fruchter}, {Goldhaber}, {Goobar}, {Gude}, {Huang}, {Ihara}, {Konishi},
  {Kowalski}, {Lidman}, {Meyers}, {Morokuma}, {Nugent}, {Perlmutter}, {Rubin},
  {Schlegel}, {Spadafora}, {Suzuki}, {Swift}, {Takanashi}, {Thomas}, \&
  {Yasuda}}]{barbary2009}
{Barbary} K. {et~al.}, 2009, \apj, 690, 1358

\bibitem[{{Berger} {et~al}\mbox{.}(2012){Berger}, {Chornock}, {Lunnan},
  {Foley}, {Czekala}, {Rest}, {Leibler}, {Soderberg}, {Roth}, {Narayan},
  {Huber}, {Milisavljevic}, {Sanders}, {Drout}, {Margutti}, {Kirshner},
  {Marion}, {Challis}, {Riess}, {Smartt}, {Burgett}, {Hodapp}, {Heasley},
  {Kaiser}, {Kudritzki}, {Magnier}, {McCrum}, {Price}, {Smith}, {Tonry}, \&
  {Wainscoat}}]{berger2012}
{Berger} E. {et~al.}, 2012, \apjl, 755, L29

\bibitem[{{Botticella} {et~al}\mbox{.}(2008){Botticella}, {Riello},
  {Cappellaro}, {Benetti}, {Altavilla}, {Pastorello}, {Turatto}, {Greggio},
  {Patat}, {Valenti}, {Zampieri}, {Harutyunyan}, {Pignata}, \&
  {Taubenberger}}]{botticella2008}
{Botticella} M.~T. {et~al.}, 2008, \aap, 479, 49

\bibitem[{{Chatzopoulos} \& {Wheeler}(2012)}]{chatzopoulos_wheeler2012}
{Chatzopoulos} E., {Wheeler} J.~C., 2012, \apj, 748, 42

\bibitem[{{Chatzopoulos} {et~al}\mbox{.}(2011){Chatzopoulos}, {Wheeler},
  {Vinko}, {Quimby}, {Robinson}, {Miller}, {Foley}, {Perley}, {Yuan},
  {Akerlof}, \& {Bloom}}]{chatzopoulos2011}
{Chatzopoulos} E. {et~al.}, 2011, \apj, 729, 143

\bibitem[{{Chevalier} \& {Irwin}(2011)}]{chevalier_irwin2011}
{Chevalier} R.~A., {Irwin} C.~M., 2011, \apjl, 729, L6

\bibitem[{{Chomiuk} {et~al}\mbox{.}(2011){Chomiuk}, {Chornock}, {Soderberg},
  {Berger}, {Chevalier}, {Foley}, {Huber}, {Narayan}, {Rest}, {Gezari},
  {Kirshner}, {Riess}, {Rodney}, {Smartt}, {Stubbs}, {Tonry}, {Wood-Vasey},
  {Burgett}, {Chambers}, {Czekala}, {Flewelling}, {Forster}, {Kaiser},
  {Kudritzki}, {Magnier}, {Martin}, {Morgan}, {Neill}, {Price}, {Roth},
  {Sanders}, \& {Wainscoat}}]{chomiuk2011}
{Chomiuk} L. {et~al.}, 2011, \apj, 743, 114

\bibitem[{{Colgate} \& {Cameron}(1963)}]{colgate_cameron1963}
{Colgate} S.~A., {Cameron} A.~G.~W., 1963, \nat, 200, 870

\bibitem[{{Cooke} {et~al}\mbox{.}(2012){Cooke}, {Sullivan}, {Gal-Yam},
  {Barton}, {Carlberg}, {Ryan-Weber}, {Horst}, {Omori}, \&
  {D{\'{\i}}az}}]{cooke2012}
{Cooke} J. {et~al.}, 2012, \nat, 491, 228

\bibitem[{{Dahlen} {et~al}\mbox{.}(2012){Dahlen}, {Strolger}, {Riess},
  {Mattila}, {Kankare}, \& {Mobasher}}]{dahlen2012}
{Dahlen} T., {Strolger} L.-G., {Riess} A.~G., {Mattila} S., {Kankare} E.,
  {Mobasher} B., 2012, \apj, 757, 70

\bibitem[{{Dessart} {et~al}\mbox{.}(2012){Dessart}, {Hillier}, {Waldman},
  {Livne}, \& {Blondin}}]{dessart2012}
{Dessart} L., {Hillier} D.~J., {Waldman} R., {Livne} E., {Blondin} S., 2012,
  ArXiv e-prints

\bibitem[{{Drake} {et~al}\mbox{.}(2011){Drake}, {Djorgovski}, {Mahabal},
  {Anderson}, {Roy}, {Mohan}, {Ravindranath}, {Frail}, {Gezari}, {Neill}, {Ho},
  {Prieto}, {Thompson}, {Thorstensen}, {Wagner}, {Kowalski}, {Chiang}, {Grove},
  {Schinzel}, {Wood}, {Carrasco}, {Recillas}, {Kewley}, {Archana}, {Basu},
  {Wadadekar}, {Kumar}, {Myers}, {Phinney}, {Williams}, {Graham}, {Catelan},
  {Beshore}, {Larson}, \& {Christensen}}]{drake2011}
{Drake} A.~J. {et~al.}, 2011, \apj, 735, 106

\bibitem[{{Drake} {et~al}\mbox{.}(2009){Drake}, {Djorgovski}, {Mahabal},
  {Beshore}, {Larson}, {Graham}, {Williams}, {Christensen}, {Catelan},
  {Boattini}, {Gibbs}, {Hill}, \& {Kowalski}}]{drake2009}
---, 2009, \apj, 696, 870

\bibitem[{{Drake} {et~al}\mbox{.}(2010){Drake}, {Djorgovski}, {Prieto},
  {Mahabal}, {Balam}, {Williams}, {Graham}, {Catelan}, {Beshore}, \&
  {Larson}}]{drake2010}
---, 2010, \apjl, 718, L127

\bibitem[{{Filippenko}(1989)}]{filippenko1989}
{Filippenko} A.~V., 1989, \aj, 97, 726

\bibitem[{{Filippenko} {et~al}\mbox{.}(2001){Filippenko}, {Li}, {Treffers}, \&
  {Modjaz}}]{filippenko2001}
{Filippenko} A.~V., {Li} W.~D., {Treffers} R.~R., {Modjaz} M., 2001, in
  Astronomical Society of the Pacific Conference Series, Vol. 246, IAU Colloq.
  183: Small Telescope Astronomy on Global Scales, {Paczynski} B., {Chen}
  W.-P., {Lemme} C., eds., p. 121

\bibitem[{{Foley} {et~al}\mbox{.}(2006){Foley}, {Li}, {Moore}, {Wong},
  {Pooley}, \& {Filippenko}}]{foley2006}
{Foley} R.~J., {Li} W., {Moore} M., {Wong} D.~S., {Pooley} D., {Filippenko}
  A.~V., 2006, Central Bureau Electronic Telegrams, 695, 1

\bibitem[{{Gal-Yam}(2012)}]{galyam2012}
{Gal-Yam} A., 2012, ArXiv e-prints

\bibitem[{{Gal-Yam} {et~al}\mbox{.}(2009){Gal-Yam}, {Mazzali}, {Ofek},
  {Nugent}, {Kulkarni}, {Kasliwal}, {Quimby}, {Filippenko}, {Cenko},
  {Chornock}, {Waldman}, {Kasen}, {Sullivan}, {Beshore}, {Drake}, {Thomas},
  {Bloom}, {Poznanski}, {Miller}, {Foley}, {Silverman}, {Arcavi}, {Ellis}, \&
  {Deng}}]{galyam2009}
{Gal-Yam} A. {et~al.}, 2009, \nat, 462, 624

\bibitem[{{Gehrels}(1986)}]{gehrels1986}
{Gehrels} N., 1986, \apj, 303, 336

\bibitem[{{Genzel} {et~al}\mbox{.}(2000){Genzel}, {Pichon}, {Eckart},
  {Gerhard}, \& {Ott}}]{genzel2000}
{Genzel} R., {Pichon} C., {Eckart} A., {Gerhard} O.~E., {Ott} T., 2000, \mnras,
  317, 348

\bibitem[{{Gezari} {et~al}\mbox{.}(2009){Gezari}, {Halpern}, {Grupe}, {Yuan},
  {Quimby}, {McKay}, {Chamarro}, {Sisson}, {Akerlof}, {Wheeler}, {Brown},
  {Cenko}, {Rau}, {Djordjevic}, \& {Terndrup}}]{gezari2009}
{Gezari} S. {et~al.}, 2009, \apj, 690, 1313

\bibitem[{{Graur} {et~al}\mbox{.}(2011){Graur}, {Poznanski}, {Maoz}, {Yasuda},
  {Totani}, {Fukugita}, {Filippenko}, {Foley}, {Silverman}, {Gal-Yam},
  {Horesh}, \& {Jannuzi}}]{graur2011}
{Graur} O. {et~al.}, 2011, \mnras, 417, 916

\bibitem[{{Greene} \& {Ho}(2007)}]{greene2007}
{Greene} J.~E., {Ho} L.~C., 2007, \apj, 667, 131

\bibitem[{{Hadjiyska} {et~al}\mbox{.}(2012){Hadjiyska}, {Rabinowitz}, {Baltay},
  {Ellman}, {Nugent}, {Zinn}, {Horowitz}, {McKinnon}, \&
  {Miller}}]{hadjiyska2012}
{Hadjiyska} E. {et~al.}, 2012, in IAU Symposium, Vol. 285, IAU Symposium,
  {Griffin} E., {Hanisch} R., {Seaman} R., eds., pp. 324--326

\bibitem[{{Harutyunyan} {et~al}\mbox{.}(2006){Harutyunyan}, {Benetti},
  {Turatto}, {Cappellaro}, {Elias-Rosa}, \& {Andreuzzi}}]{harutyunyan2006}
{Harutyunyan} A., {Benetti} S., {Turatto} M., {Cappellaro} E., {Elias-Rosa} N.,
  {Andreuzzi} G., 2006, Central Bureau Electronic Telegrams, 647, 1

\bibitem[{{Hatano}, {Branch} \& {Deaton}(1998){Hatano}, {Branch}, \&
  {Deaton}}]{hatano1998}
{Hatano} K., {Branch} D., {Deaton} J., 1998, \apj, 502, 177

\bibitem[{{Hatano} {et~al}\mbox{.}(1999){Hatano}, {Branch}, {Fisher},
  {Millard}, \& {Baron}}]{hatano1999}
{Hatano} K., {Branch} D., {Fisher} A., {Millard} J., {Baron} E., 1999, \apjs,
  121, 233

\bibitem[{{Hopkins} \& {Beacom}(2006)}]{hopkins_beacom2006}
{Hopkins} A.~M., {Beacom} J.~F., 2006, \apj, 651, 142

\bibitem[{{Horiuchi} {et~al}\mbox{.}(2011){Horiuchi}, {Beacom}, {Kochanek},
  {Prieto}, {Stanek}, \& {Thompson}}]{horiuchi2011}
{Horiuchi} S., {Beacom} J.~F., {Kochanek} C.~S., {Prieto} J.~L., {Stanek}
  K.~Z., {Thompson} T.~A., 2011, \apj, 738, 154

\bibitem[{{Kaiser} {et~al}\mbox{.}(2010){Kaiser}, {Burgett}, {Chambers},
  {Denneau}, {Heasley}, {Jedicke}, {Magnier}, {Morgan}, {Onaka}, \&
  {Tonry}}]{kaiser2010}
{Kaiser} N. {et~al.}, 2010, in Society of Photo-Optical Instrumentation
  Engineers (SPIE) Conference Series, Vol. 7733, Society of Photo-Optical
  Instrumentation Engineers (SPIE) Conference Series

\bibitem[{{Kasen} \& {Bildsten}(2010)}]{kasen_bildsten2010}
{Kasen} D., {Bildsten} L., 2010, \apj, 717, 245

\bibitem[{{Kelly}(2007)}]{kelly2007}
{Kelly} B.~C., 2007, \apj, 665, 1489

\bibitem[{{Knop} {et~al}\mbox{.}(1999){Knop}, {Aldering}, {Deustua},
  {Goldhaber}, {Kim}, {Nugent}, {Helin}, {Pravdo}, {Rabinowitz}, \&
  {Lawrence}}]{knop1999}
{Knop} R. {et~al.}, 1999, \iaucirc, 7128, 1

\bibitem[{{Lauer} {et~al}\mbox{.}(2012){Lauer}, {Bender}, {Kormendy},
  {Rosenfield}, \& {Green}}]{lauer2012}
{Lauer} T.~R., {Bender} R., {Kormendy} J., {Rosenfield} P., {Green} R.~F.,
  2012, \apj, 745, 121

\bibitem[{{Law} {et~al}\mbox{.}(2009){Law}, {Kulkarni}, {Dekany}, {Ofek},
  {Quimby}, {Nugent}, {Surace}, {Grillmair}, {Bloom}, {Kasliwal}, {Bildsten},
  {Brown}, {Cenko}, {Ciardi}, {Croner}, {Djorgovski}, {van Eyken},
  {Filippenko}, {Fox}, {Gal-Yam}, {Hale}, {Hamam}, {Helou}, {Henning},
  {Howell}, {Jacobsen}, {Laher}, {Mattingly}, {McKenna}, {Pickles},
  {Poznanski}, {Rahmer}, {Rau}, {Rosing}, {Shara}, {Smith}, {Starr},
  {Sullivan}, {Velur}, {Walters}, \& {Zolkower}}]{law2009}
{Law} N.~M. {et~al.}, 2009, \pasp, 121, 1395

\bibitem[{{Leloudas} {et~al}\mbox{.}(2012){Leloudas}, {Chatzopoulos}, {Dilday},
  {Gorosabel}, {Vinko}, {Gallazzi}, {Wheeler}, {Bassett}, {Fischer}, {Frieman},
  {Fynbo}, {Goobar}, {Jel{\'{\i}}nek}, {Malesani}, {Nichol}, {Nordin},
  {{\"O}stman}, {Sako}, {Schneider}, {Smith}, {Sollerman}, {Stritzinger},
  {Th{\"o}ne}, \& {de Ugarte Postigo}}]{leloudas2012}
{Leloudas} G. {et~al.}, 2012, \aap, 541, A129

\bibitem[{{Levin} \& {Beloborodov}(2003)}]{levin2003}
{Levin} Y., {Beloborodov} A.~M., 2003, \apjl, 590, L33

\bibitem[{{Li} {et~al}\mbox{.}(2011){Li}, {Leaman}, {Chornock}, {Filippenko},
  {Poznanski}, {Ganeshalingam}, {Wang}, {Modjaz}, {Jha}, {Foley}, \&
  {Smith}}]{li2011a}
{Li} W. {et~al.}, 2011, \mnras, 412, 1441

\bibitem[{{Miller} {et~al}\mbox{.}(2009){Miller}, {Chornock}, {Perley},
  {Ganeshalingam}, {Li}, {Butler}, {Bloom}, {Smith}, {Modjaz}, {Poznanski},
  {Filippenko}, {Griffith}, {Shiode}, \& {Silverman}}]{miller2009}
{Miller} A.~A. {et~al.}, 2009, \apj, 690, 1303

\bibitem[{{Modjaz} {et~al}\mbox{.}(2009){Modjaz}, {Li}, {Butler}, {Chornock},
  {Perley}, {Blondin}, {Bloom}, {Filippenko}, {Kirshner}, {Kocevski},
  {Poznanski}, {Hicken}, {Foley}, {Stringfellow}, {Berlind}, {Barrado y
  Navascues}, {Blake}, {Bouy}, {Brown}, {Challis}, {Chen}, {de Vries},
  {Dufour}, {Falco}, {Friedman}, {Ganeshalingam}, {Garnavich}, {Holden},
  {Illingworth}, {Lee}, {Liebert}, {Marion}, {Olivier}, {Prochaska},
  {Silverman}, {Smith}, {Starr}, {Steele}, {Stockton}, {Williams}, \&
  {Wood-Vasey}}]{modjaz2009}
{Modjaz} M. {et~al.}, 2009, \apj, 702, 226

\bibitem[{{Moriya} \& {Tominaga}(2012)}]{moriya_tominaga2012}
{Moriya} T.~J., {Tominaga} N., 2012, \apj, 747, 118

\bibitem[{{Neill} {et~al}\mbox{.}(2011){Neill}, {Sullivan}, {Gal-Yam},
  {Quimby}, {Ofek}, {Wyder}, {Howell}, {Nugent}, {Seibert}, {Martin},
  {Overzier}, {Barlow}, {Foster}, {Friedman}, {Morrissey}, {Neff},
  {Schiminovich}, {Bianchi}, {Donas}, {Heckman}, {Lee}, {Madore}, {Milliard},
  {Rich}, \& {Szalay}}]{neill2011}
{Neill} J.~D. {et~al.}, 2011, \apj, 727, 15

\bibitem[{{Ofek} {et~al}\mbox{.}(2007){Ofek}, {Cameron}, {Kasliwal}, {Gal-Yam},
  {Rau}, {Kulkarni}, {Frail}, {Chandra}, {Cenko}, {Soderberg}, \&
  {Immler}}]{ofek2007}
{Ofek} E.~O. {et~al.}, 2007, \apjl, 659, L13

\bibitem[{{Ouyed} {et~al}\mbox{.}(2012){Ouyed}, {Kostka}, {Koning}, {Leahy}, \&
  {Steffen}}]{ouyed2012}
{Ouyed} R., {Kostka} M., {Koning} N., {Leahy} D.~A., {Steffen} W., 2012,
  \mnras, 423, 1652

\bibitem[{{Pastorello} {et~al}\mbox{.}(2010){Pastorello}, {Smartt},
  {Botticella}, {Maguire}, {Fraser}, {Smith}, {Kotak}, {Magill}, {Valenti},
  {Young}, {Gezari}, {Bresolin}, {Kudritzki}, {Howell}, {Rest}, {Metcalfe},
  {Mattila}, {Kankare}, {Huang}, {Urata}, {Burgett}, {Chambers}, {Dombeck},
  {Flewelling}, {Grav}, {Heasley}, {Hodapp}, {Kaiser}, {Luppino}, {Lupton},
  {Magnier}, {Monet}, {Morgan}, {Onaka}, {Price}, {Rhoads}, {Siegmund},
  {Stubbs}, {Sweeney}, {Tonry}, {Wainscoat}, {Waterson}, {Waters}, \&
  {Wynn-Williams}}]{pastorello2010}
{Pastorello} A. {et~al.}, 2010, \apjl, 724, L16

\bibitem[{{Prieto} {et~al}\mbox{.}(2006){Prieto}, {Garnavich}, {Chronister}, \&
  {Connick}}]{prieto2006}
{Prieto} J.~L., {Garnavich} P., {Chronister} A., {Connick} P., 2006, Central
  Bureau Electronic Telegrams, 648, 1

\bibitem[{{Quimby}(2006)}]{quimby_phd}
{Quimby} R.~M., 2006, PhD thesis, The University of Texas at Austin

\bibitem[{{Quimby}(2008)}]{quimby2008}
---, 2008, in Astronomical Society of the Pacific Conference Series, Vol. 393,
  New Horizons in Astronomy, {Frebel} A., {Maund} J.~R., {Shen} J., {Siegel}
  M.~H., eds., p. 141

\bibitem[{{Quimby} {et~al}\mbox{.}(2007){Quimby}, {Aldering}, {Wheeler},
  {H{\"o}flich}, {Akerlof}, \& {Rykoff}}]{quimby2007c}
{Quimby} R.~M., {Aldering} G., {Wheeler} J.~C., {H{\"o}flich} P., {Akerlof}
  C.~W., {Rykoff} E.~S., 2007, \apjl, 668, L99

\bibitem[{{Quimby} {et~al}\mbox{.}(2011){Quimby}, {Kulkarni}, {Kasliwal},
  {Gal-Yam}, {Arcavi}, {Sullivan}, {Nugent}, {Thomas}, {Howell}, {Nakar},
  {Bildsten}, {Theissen}, {Law}, {Dekany}, {Rahmer}, {Hale}, {Smith}, {Ofek},
  {Zolkower}, {Velur}, {Walters}, {Henning}, {Bui}, {McKenna}, {Poznanski},
  {Cenko}, \& {Levitan}}]{quimby2011}
{Quimby} R.~M. {et~al.}, 2011, \nat, 474, 487

\bibitem[{{Quimby} {et~al}\mbox{.}(2012){Quimby}, {Yuan}, {Akerlof}, {Wheeler},
  \& {Warren}}]{quimby2012}
{Quimby} R.~M., {Yuan} F., {Akerlof} C., {Wheeler} J.~C., {Warren} M.~S., 2012,
  ArXiv e-prints

\bibitem[{{Rau} {et~al}\mbox{.}(2009){Rau}, {Kulkarni}, {Law}, {Bloom},
  {Ciardi}, {Djorgovski}, {Fox}, {Gal-Yam}, {Grillmair}, {Kasliwal}, {Nugent},
  {Ofek}, {Quimby}, {Reach}, {Shara}, {Bildsten}, {Cenko}, {Drake},
  {Filippenko}, {Helfand}, {Helou}, {Howell}, {Poznanski}, \&
  {Sullivan}}]{rau2009}
{Rau} A. {et~al.}, 2009, \pasp, 121, 1334

\bibitem[{{Rest} {et~al}\mbox{.}(2011){Rest}, {Foley}, {Gezari}, {Narayan},
  {Draine}, {Olsen}, {Huber}, {Matheson}, {Garg}, {Welch}, {Becker}, {Challis},
  {Clocchiatti}, {Cook}, {Damke}, {Meixner}, {Miknaitis}, {Minniti}, {Morelli},
  {Nikolaev}, {Pignata}, {Prieto}, {Smith}, {Stubbs}, {Suntzeff}, {Walker},
  {Wood-Vasey}, {Zenteno}, {Wyrzykowski}, {Udalski}, {Szyma{\'n}ski}, {Kubiak},
  {Pietrzy{\'n}ski}, {Soszy{\'n}ski}, {Szewczyk}, {Ulaczyk}, \&
  {Poleski}}]{rest2011}
{Rest} A. {et~al.}, 2011, \apj, 729, 88

\bibitem[{{Scalzo} {et~al}\mbox{.}(2010){Scalzo}, {Aldering}, {Antilogus},
  {Aragon}, {Bailey}, {Baltay}, {Bongard}, {Buton}, {Childress}, {Chotard},
  {Copin}, {Fakhouri}, {Gal-Yam}, {Gangler}, {Hoyer}, {Kasliwal}, {Loken},
  {Nugent}, {Pain}, {P{\'e}contal}, {Pereira}, {Perlmutter}, {Rabinowitz},
  {Rau}, {Rigaudier}, {Runge}, {Smadja}, {Tao}, {Th\ omas}, {Weaver}, \&
  {Wu}}]{scalzo2010}
{Scalzo} R.~A. {et~al.}, 2010, \apj, 713, 1073

\bibitem[{{Schlegel}, {Finkbeiner} \& {Davis}(1998){Schlegel}, {Finkbeiner}, \&
  {Davis}}]{sfd1998}
{Schlegel} D.~J., {Finkbeiner} D.~P., {Davis} M., 1998, \apj, 500, 525

\bibitem[{{Smith} {et~al}\mbox{.}(2008{\natexlab{a}}){Smith}, {Chornock}, {Li},
  {Ganeshalingam}, {Silverman}, {Foley}, {Filippenko}, \& {Barth}}]{smith2008}
{Smith} N., {Chornock} R., {Li} W., {Ganeshalingam} M., {Silverman} J.~M.,
  {Foley} R.~J., {Filippenko} A.~V., {Barth} A.~J., 2008{\natexlab{a}}, \apj,
  686, 467

\bibitem[{{Smith} {et~al}\mbox{.}(2008{\natexlab{b}}){Smith}, {Foley}, {Bloom},
  {Li}, {Filippenko}, {Gavazzi}, {Ghez}, {Konopacky}, {Malkan}, {Marshall},
  {Pooley}, {Treu}, \& {Woo}}]{smith2008b}
{Smith} N. {et~al.}, 2008{\natexlab{b}}, \apj, 686, 485

\bibitem[{{Smith} {et~al}\mbox{.}(2007){Smith}, {Li}, {Foley}, {Wheeler},
  {Pooley}, {Chornock}, {Filippenko}, {Silverman}, {Quimby}, {Bloom}, \&
  {Hansen}}]{smith2007}
---, 2007, \apj, 666, 1116

\bibitem[{{Smith} \& {McCray}(2007)}]{smith_mccray2007}
{Smith} N., {McCray} R., 2007, \apjl, 671, L17

\bibitem[{{Soderberg} {et~al}\mbox{.}(2006){Soderberg}, {Kulkarni}, {Nakar},
  {Berger}, {Cameron}, {Fox}, {Frail}, {Gal-Yam}, {Sari}, {Cenko}, {Kasliwal},
  {Chevalier}, {Piran}, {Price}, {Schmidt}, {Pooley}, {Moon}, {Penprase},
  {Ofek}, {Rau}, {Gehrels}, {Nousek}, {Burrows}, {Persson}, \&
  {McCarthy}}]{soderberg2006}
{Soderberg} A.~M. {et~al.}, 2006, \nat, 442, 1014

\bibitem[{{Tanaka} {et~al}\mbox{.}(2012){Tanaka}, {Moriya}, {Yoshida}, \&
  {Nomoto}}]{tanaka2012}
{Tanaka} M., {Moriya} T.~J., {Yoshida} N., {Nomoto} K., 2012, \mnras, 422, 2675

\bibitem[{{Vinko} {et~al}\mbox{.}(2010){Vinko}, {Zheng}, {Romadan}, {Quimby},
  {Whallon}, {Pandey}, {Fang}, {Akerlof}, {Pasque}, {Verkinderen}, {Wheeler},
  {Chatzopoulos}, \& {Caldwell}}]{vinko2010}
{Vinko} J. {et~al.}, 2010, Central Bureau Electronic Telegrams, 2556, 1

\bibitem[{{Woosley}(2010)}]{woosley2010}
{Woosley} S.~E., 2010, \apjl, 719, L204

\bibitem[{{Yoon}, {Dierks} \& {Langer}(2012){Yoon}, {Dierks}, \&
  {Langer}}]{yoon_langer2012}
{Yoon} S.-C., {Dierks} A., {Langer} N., 2012, \aap, 542, A113

\bibitem[{{Yuan}(2010)}]{yuan_phd}
{Yuan} F., 2010, PhD thesis, University of Michigan

\bibitem[{{Yuan} {et~al}\mbox{.}(2010){Yuan}, {Quimby}, {Wheeler}, {Vink{\'o}},
  {Chatzopoulos}, {Akerlof}, {Kulkarni}, {Miller}, {\ McKay}, \&
  {Aharonian}}]{yuan2010}
{Yuan} F. {et~al.}, 2010, \apj, 715, 1338

\bibitem[{{Y{\"u}ksel} {et~al}\mbox{.}(2008){Y{\"u}ksel}, {Kistler}, {Beacom},
  \& {Hopkins}}]{yuksel2008}
{Y{\"u}ksel} H., {Kistler} M.~D., {Beacom} J.~F., {Hopkins} A.~M., 2008, \apjl,
  683, L5

\bibitem[{{Zwicky}(1938)}]{zwicky1938}
{Zwicky} F., 1938, \apj, 88, 529

\end{thebibliography}
\end{document}